\documentclass[12pt]{article}
\DeclareUnicodeCharacter{FF0C}{,}
\usepackage{graphicx} 
\usepackage{bm}
\usepackage{verbatim, color, array}

\usepackage{amsthm, amssymb}
\usepackage{float}
\usepackage{amsmath}
\usepackage{hhline}
\usepackage{booktabs}
\usepackage{soul}
\usepackage[super]{natbib}
\usepackage[table]{xcolor} 
\usepackage{hyperref}
\usepackage{multirow}
\usepackage{nicematrix}
\usepackage{threeparttable}
\usepackage{enumitem}
\usepackage{tikz}
\usepackage{appendix}
\usetikzlibrary{shapes, arrows, positioning, calc, shadows, fit}
\usepackage{longtable}
\usepackage[affil-it]{authblk} 
\usepackage{ulem}
\usepackage{threeparttable}
\title{TITE-STEIN: Time-to-event Simple Toxicity and Efficacy Interval Design to Accelerate Phase I/II Trials}
\date{}
\author[1]{Hao Sun}
\author[2]{Jieqi Tu}
\author[1]{Revathi Ananthakrishnan}
\author[3]{Eunhee Kim \thanks{Corresponding author: *Eunhee Kim, Biometrics, Nurix Therapeutics, California, USA. Email: \href{mailto:ekim@nurixtx.com}{ekim@nurixtx.com}}}
\affil[1]{Global Biometrics \& Data Sciences, Bristol Myers Squibb, New Jersey, USA}
\affil[2]{Division of Epidemiology and Biostatistics, School of Public Health, University of Illinois at Chicago, Illinois, USA}
\affil[3]{Biometrics, Nurix Therapeutics, California, USA}

\begin{document}
\maketitle

\begin{abstract}
Oncology dose-finding trials are shifting from identifying the maximum tolerated dose (MTD) to determining the optimal biological dose (OBD), driven by the need for efficient methods that consider both toxicity and efficacy. This is particularly important for novel therapies, such as immunotherapies and molecularly targeted therapies, which often exhibit non-monotonic dose-efficacy curves. However, making timely adaptive dosing decisions is challenging due to the rapid patient accrual rate and the late-onset toxicity and/or efficacy outcomes associated with these therapies. The Simple Toxicity and Efficacy Interval (STEIN) design has demonstrated strong performance in accommodating diverse dose-efficacy patterns and incorporating both toxicity and efficacy outcomes to select the OBD. However, the rapid accrual of patients and the often-delayed onset of toxicity and/or efficacy pose challenges to timely adaptive dose decisions. To address these challenges, we propose TITE-STEIN, a model-assisted design that incorporates time-to-event (TITE) outcomes for toxicity and/or efficacy, by extending STEIN. In this article, we demonstrate that TITE-STEIN significantly shortens trial duration compared to STEIN. Furthermore, by integrating an OBD verification procedure during OBD selection, TITE-STEIN effectively mitigates the risk of exposing patients to inadmissible doses when the OBD does not exist. Extensive simulations demonstrate that TITE-STEIN outperforms existing TITE designs, including TITE-BONI12, TITE-BOIN-ET, LO-TC, and Joint TITE-CRM, by selecting the OBD more accurately, allocating more patients to it, and improving overdose control.

\end{abstract}


\newpage

\section{Introduction}
Traditional Phase I oncology dose-finding trials primarily aim to identify the maximum tolerated dose (MTD), which is the highest dose with a probability of dose-limiting toxicity (DLT) below the target toxicity rate. The dose-finding designs used to determine the MTD can be classified into rule-based, model-based, and model-assisted designs. The rule-based 3+3 design \cite{storer1989design} is commonly used due to its simplicity. However, the 3+3 design is often criticized for allocating significantly fewer patients to the true MTD, assigning a higher number of patients to lower dose levels, and rarely exploring doses above the target DLT, compared to other designs discussed below \cite{chiuzan20243+}. In contrast, model-based designs, such as the continual reassessment method (CRM)\cite{CRM1990}, dose escalation with overdose control (EWOC) \cite{Babb1998}, and the Bayesian logistic regression model (BLRM) \cite{neuenschwander2008critical}, leverage statistical models to estimate the dose-toxicity relationship and identify the MTD, thereby improving trial efficiency and the accuracy of MTD identification. Model-assisted designs, such as the Bayesian optimal interval (BOIN) design\cite{liu2015bayesian}, the modified toxicity probability interval (mTPI) design\cite{ji2013modified}, and the mTPI-2 design\cite{yan2017keyboard}, strike a balance between model complexity and ease of practical implementation. These designs to identify the MTD assume that efficacy increases with dose, and the MTD is often considered the dose with the highest efficacy among tolerable dose levels.

The FDA's Optimus Project advocates for dose optimization to identify safe and efficacious doses\cite{fda_project_optimus}. As a result, the focus of dose-finding trials has shifted from determining the MTD to identifying the optimal biological dose (OBD) that maximizes the risk-benefit trade-off by evaluating toxicity and efficacy simultaneously. Several model-based designs have been developed to jointly model the dose-toxicity and dose-efficacy curves, such as Eff-Tox \cite{thall2004dose}, the B-dynamic model\cite{liu2016robust}, and EWOUC-NETS\cite{tu2024bayesian}. However, for immuno-oncology drugs and molecularly targeted therapies, where efficacy does not always increase with dose, model-assisted designs are often preferred over model-based designs as the model-assisted designs generally do not require model assumptions about the dose-efficacy curves. To determine the OBD while accounting for uncertainty in the dose-efficacy curve, numerous model-assisted dose-finding designs have been developed. Examples include mTPI-based designs such as TEPI-2 \cite{li2020tepi}, PRINTE \cite{lin2021probability}, and uTPI \cite{shi2021utpi}, and BOIN-based designs such as BOIN12 \cite{Lin2020}, BOIN-ET \cite{Takeda2018}, and STEIN\cite{lin2017stein}. 


Escalating resource requirements for cancer clinical trials, encompassing patient recruitment, data management infrastructure, and individual treatment costs, underscores the critical need for improved trial efficiency and accelerated patient enrollment \cite{markey2024clinical, sertkaya2024costs}. However, making timely adaptive dosing decisions remains challenging, particularly due to the rapid accrual and late-onset outcomes of toxicity and efficacy that are common in trials of novel molecularly targeted agents and immunotherapies. These factors can delay dosing decisions for subsequent cohorts, as outcomes from earlier patients may still be pending\cite{Yuan2008}. To address these issues and accelerate early-phase clinical trials, many dose-finding designs incorporating time-to-event (TITE) outcomes for toxicity and/or efficacy have been proposed. Model-based TITE designs include TITE-CRM\cite{Cheung2000} and EWOC-NETS-TITE \cite{Chen2014} for MTD identification, as well as LO-TC\cite{jin2014using} and joint TITE-CRM \cite{barnett2024joint} for OBD identification. Model-assisted TITE designs include TITE-BOIN \cite{yuan2018time} and TITE-keyboard \cite{lin2020time} for MTD identification, along with TITE-BOIN12 \cite{zhou2022tite} and TITE-BOIN-ET \cite{takeda2020tite} for OBD identification. These TITE designs have demonstrated improved trial efficiency while maintaining accuracy in MTD/OBD selection. A key advantage of TITE designs is that they allow the enrollment of the next cohort of patients even without obtaining complete data from the currently enrolled patients, using partial outcome information to make real-time dosing decisions. 

Our work is motivated by the results of \citet{sun2024statistical}, who comprehensively compared the performance of seven model-assisted OBD-finding designs, namely BOIN-ET, BOIN12, TEPI-2, UBI, PRINTE, STEIN, and uTPI. Their findings indicated that STEIN consistently demonstrated high accuracy in selecting the true OBD across various dose-efficacy patterns, with the best overdose control when the OBD existed. Additionally, PRINTE had the highest probability of not selecting any dose when the OBD did not exist, owing to its implementation of an additional step for evaluating the appropriateness of the selected dose. 

We introduce TITE-STEIN, a novel model-assisted TITE design for identifying the optimal dose that considers both toxicity and efficacy outcomes.  Leveraging Sun et al.'s findings on STEIN and PRINTE\cite{sun2024statistical}, TITE-STEIN extends STEIN and incorporates PRINTE's OBD verification procedure into its OBD selection. Moreover, TITE-STEIN addresses the growing need to accommodate fast patient accrual rates and late-onset toxicity and efficacy. For various dose-efficacy relationships, we compare TITE-STEIN with STEIN and four TITE designs, namely TITE-BOIN12, TITE-BOIN-ET, LO-TC, and joint TITE-CRM, via simulation studies and a case study using data from a trial.

TITE-STEIN demonstrated similar accuracy of OBD selection and percentage of patient allocation to the OBD compared to STEIN when the OBD existed, but with shorter trial durations, and better performance in terminating trials when no OBD existed. Furthermore, TITE-STEIN exhibited a higher probability of selecting the true OBD and assigned more patients to it across a broad range of scenarios compared to the other TITE designs considered.

The rest of the paper is organized as follows: Section~\ref{sec2:method} details the methodology of TITE-STEIN. Section~\ref{sec3:simu} presents extensive simulation studies and sensitivity analyses to compare TITE-STEIN with STEIN, TITE-BOIN12, TITE-BOIN-ET, LO-TC, and Joint TITE-CRM. Section~\ref{sec4:case} demonstrates an application of the TITE-STEIN design through a case study, compared with TITE-BOIN12 and TITE-BOIN-ET. Section~\ref{sec5:software} provides software information, and Section~\ref {sec6:discussion} concludes with a discussion.

\section{Method}\label{sec2:method}
\subsection{Notations}
Consider $D$ dose levels in an early-phase clinical trial. Let $Y_T$ and $Y_E$ represent binary outcomes for toxicity and efficacy, where $Y_T = 1$ indicates a DLT and $Y_E = 1$ denotes an efficacy response. We assume that $Y_T$ and $Y_E$ are independent. Define $n_d$ as the total number of patients assigned to dose level $d$. Let $n_{d,T}$ represent the number of patients experiencing a DLT and $n_{d,E}$ denote the number of patients who exhibit an efficacy response at dose level $d$. Let $p_T$ and $q_E$ be the maximum acceptable toxicity probability and the minimum acceptable efficacy probability, respectively. The true probabilities of toxicity and efficacy for dose level $d$ are denoted by $p_d$ and $q_d$ for $d = 1, \ldots, D$. As is common in most dose-finding trials, we assume that $p_d$ increases with dose level, such that $p_1 < \ldots < p_D$, while this assumption may not hold for the efficacy outcome $q_d$. The observed toxicity and response probabilities are given by $\hat{p}_d = n_{d, T}/n_d$ and $\hat{q}_d = n_{d,E}/n_d$, respectively.

A dose level $d_{MTD}$ is defined as the MTD if it has the highest acceptable toxicity probability, i.e., $d_{MTD} = \arg\max_{d} p_d I(p_d \leq p_T)$, where $I(\cdot)$ is an indicator function. If no dose level has an acceptable toxicity probability, $d_{MTD}$ does not exist. A dose level $d$ is called an admissible dose (AD) if it is not higher than $d_{MTD}$ with an acceptable efficacy probability. In other words, $d$ is an AD if $p_d \leq p_T$ and $q_d \geq q_E$. 

If at least one dose is an AD, the OBD must exist. However, the selection criteria for the OBD can vary across different designs. For instance, BOIN-ET\cite{Takeda2018} and TITE-BOIN-ET \cite{takeda2020tite} select the dose with the highest efficacy probability as the OBD among the ADs, i.e., $d_{OBD} = \arg\max_d q_d I(p_d\leq p_T, q_d\geq q_E)$. A more general approach uses a utility function $U(p_d, q_d)$ that incorporates both toxicity and efficacy probabilities to balance the benefit-risk trade-off. Here, the OBD is defined as the dose level with the highest utility function value, i.e., $d_{OBD} = \arg\max_d U(p_d, q_d)I(p_d \leq p_T, q_d \geq q_E)$. For example, STEIN \cite{lin2017stein} uses a linear function $U(p_d, q_d) = q_d - w_1 p_d - w_2 p_dI(p_d > \phi)$ for OBD selection, while Eff-Tox \cite{thall2004dose} and LO-TC \cite{jin2014using} apply a quadratic function $U(p_d, q_d) = \beta_1q_d + \beta_2q_d^2 + p_d$. In addition, BOIN12\cite{lin2020boin12} and TITE-BOIN12 \cite{zhou2022tite} define another utility function $U(p_d, q_d) = \sum_{a=0}^1\sum_{b=0}^1 u_{ab}p_d^{a}(1-p_d)^{1-a}q_d^{b}(1-q_d)^{1-b}$ based on four different combinations of binary toxicity and efficacy outcomes, where $\{u_{11}, u_{10}, u_{01}, u_{00}\}$ are called utility scores. Users can adjust the parameters in the various utility functions to reflect their specific preferences in the toxicity-efficacy trade-off.

\subsection{STEIN Design}
STEIN is a model-assisted design for OBD selection that utilizes a frequentist model-averaging technique to estimate the efficacy probabilities \cite{lin2017stein}. Let $\phi$ denote the target toxicity probability pre-specified by clinicians. As an extension of BOIN, STEIN applies two cut points, $\phi_1$ and $\phi_2$, on the toxicity interval, with $0 < \phi_1 < \phi < \phi_2 < 1$. The two cutoffs $\phi_1$ and $\phi_2$ split the toxicity interval into three intervals: the toxicity probability $p_d$ that falls into $\Phi_1 = [0, \phi_1]$, $\Phi_2 = (\phi_1, \phi_2)$, or $\Phi_3 = [\phi_2, 1]$ is classified as overly safe, acceptable, and toxic, respectively. Following \citet{liu2015bayesian}, STEIN considers three hypotheses for the toxicity probability: 
$$
H_{1d}^T: p_d = \phi_1 \quad \text{versus} \quad  H_{2d}^T: p_d = \phi \quad \text{versus} \quad  H_{3d}^T: p_d = \phi_2
$$
The probability of incorrect classification for the toxicity outcome, given the observed data, is 
$$
\Pr(H_{1d}^T)\Pr(\hat{p}_d > \phi_L\mid H_{1d}^T) + \Pr(H_{2d}^T)\Pr(\hat{p}_d \leq \phi_L \text{ or } \hat{p}_d \geq \phi_U\mid H_{2d}^T) + \Pr(H_{3d}^T)\Pr(\hat{p}_d < \phi_U\mid H_{3d}^T)
$$
where $\phi_L, \phi_U \in [0, 1]$ are two design parameters that define an indifferent tolerance interval $(\phi_L, \phi_U)$ of $\phi$. We assume a uniform prior on the hypotheses, i.e., $\Pr(H_{1d}^T) = \Pr(H_{2d}^T) = \Pr(H_{3d}^T) = 1/3$. By minimizing the probability of incorrect classification, we have
$$
\phi_L = \frac{\log(\frac{1-\phi_1}{1-\phi})}{\log[\frac{\phi(1-\phi_1)}{\phi_1(1-\phi)}]},  \quad \phi_U = \frac{\log(\frac{1-\phi}{1-\phi_2})}{\log[\frac{\phi_2(1-\phi)}{\phi(1-\phi_2)}]}. 
$$

Similarly, for the efficacy outcome, STEIN considers two hypotheses: $H_{1d}^E: q_d = \psi_1$ versus $H_{1d}^E: q_d = \psi_2$. The probability of incorrect classification for the efficacy outcome given the observed data is 
$$
\Pr(H_{1d}^E)\Pr(\hat{q}_j \geq \psi\mid H_{1d}^E) + \Pr(H_{2d}^E)\Pr(\hat{q}_j < \psi\mid H_{2d}^E), 
$$
where $\phi$ is a design parameter. With the assumption of a uniform prior, $\psi = \log(\frac{1-\psi_1}{1-\psi_2})/\log(\frac{\psi_2(1-\psi_1)}{\psi_1(1-\psi_2)})$ minimizes the probability of incorrect classification for the efficacy outcome. 

The three design parameters $\{\phi_L, \phi_U, \psi\}$ split the combination square of toxicity and efficacy intervals into three pieces: (1) promising region: $\{(\hat{p}_d, \hat{q}_d): \hat{p}_d < \phi_U, \hat{q}_d \geq \psi\}$, where the current dose is considered safe and efficacious, prompting an escalation to a higher dose level; (2) exploratory region: $\{(\hat{p}_d, \hat{q}_d): \hat{p}_d < \phi_U, \hat{q}_d < \psi\}$, where decision-making is uncertain, STEIN evaluates the admissible dose set, including the current dose, and selects the dose with the highest posterior probability of $\Pr(q_d > \psi\mid \text{Data})$ for the next patient cohort; and (3) inadmissible region: $\{(\hat{p}_d, \hat{q}_d): \hat{p}_d \geq \phi_U, \hat{q}_d \in [0, 1]\}$, indicating a toxic dose level, suggesting that dose escalation should not occur. For further details on the STEIN design algorithm, refer to \citet{lin2017stein}.

\subsection{TITE-STEIN Design}\label{TITE-STEIN}
In the proposed TITE-STEIN design, we incorporate time-to-event toxicity and efficacy outcomes to enable real-time dosing decisions, even when some toxicity or efficacy outcomes are still pending. Let $A_T$ and $A_E$ be the toxicity and efficacy assessment time windows, which can be of different lengths. Denote $t$ as the time for the next dose assignment. For the $j$-th patient, let $Y_{j,T}$ and $Y_{j, E}$ be the binary toxicity and efficacy outcomes within the assessment periods, and two dummy variables $\delta_{j, T}^{(t)}$ and $\delta_{j, E}^{(t)}$ indicate whether $Y_{j, T}$ and $Y_{j, E}$ have been ascertained ($\delta_{j, T}^{(t)}, \delta_{j, E}^{(t)}) = 1$ or are still pending ($\delta_{j, T}^{(t)}, \delta_{j, E}^{(t)}) = 0$ at time $t$. In other words, the toxicity outcome $Y_{j,T}$ is ascertained at time $t$ if patient $j$ either experiences a DLT or completes the assessment window without a DLT by time $t$. Otherwise, the toxicity outcome is pending at time $t$. A similar case holds for the efficacy assessment $Y_{j,E}$. Let $t_j$ denote the actual follow-up time for patient $j$ at time $t$, and $\{w_{j,T}, w_{j,E}\}$ be the conditional probabilities of observing the toxicity and efficacy outcomes by $t_j$. These are given as 
$$
w_{j, T}^{(t)} = \Pr(X_{j,T} \leq t_j \mid Y_{j,T} = 1), \text{ and } w_{j, E}^{(t)} = \Pr(X_{j,E} \leq t_j \mid Y_{j, E} = 1), 
$$
where $X_{j,T}$ is the time to toxicity outcome given $Y_{j, T} = 1$, and $X_{j,T}$ is the time to efficacy outcome given $Y_{j, E} = 1$. Following Lin and Yuan \cite{yuan2018time}, we assume that $X_{j,T} \sim U(0, A_T)$ and $X_{j,E} \sim U(0, A_E)$, which yields $w_{j, T}^{(t)} = \min\{t_j/A_T, 1\}$ and $w_{j,E}^{(t)} = \min\{t_j/A_E, 1\}$. 

At time $t$, denote $I_d^{(t)}$ as the set of all patients assigned to dose level $d$, and $\Theta_{d,T}^{(t)}$ as the observed interim toxicity data set at dose level $d$. For simplicity, we omit the superscripts $(t)$ of the variables $\{w_{j,T}^{(t)}, w_{j,E}^{(t)}, \delta_{j,T}^{(t)}, \delta_{j,E}^{(t)}\}$ in the following likelihood expressions. The joint likelihood of the toxicity probability $p_d$ is given by
$$\begin{aligned}
L(p_d\mid \Theta_{d,T}) & \propto \prod_{j \in I_d} p_d^{\delta_{j,T}y_{j,T}}(1-p_d)^{\delta_{j,T}(1-y_{j,T})}(1-w_{j,T}p_d)^{1-\delta_{j,T}} \\
& \approx \prod_{j \in I_d} p_d^{\delta_{j,T}y_{j,T}}(1-p_d)^{\delta_{j,T}(1-y_{j,T})}(1-p_d)^{w_{j,T}(1-\delta_{j,T})}\\
& = p_d^{n_{d,T}^{(t)}}(1-p_{d})^{\tilde{m}_{d,T}^{(t)}},
\end{aligned}$$
where $n_{d,T}^{(t)} = \sum_{j\in I_d} \delta_{j,T}^{(t)}y_{j,T}$ represents the number of patients who have experienced DLTs in $I_d^{(t)}$, $\tilde{m}_{d,T}^{(t)} = m_{d,T}^{(t)} +  \sum_{j\in I_d} (1-\delta_{j,T}^{(t)})w_{j,T}^{(t)}$ is the effective number of patients who have not experienced toxicity, and $m_{d,T}^{(t)} = \sum_{j\in I_d} \delta_{j,T}^{(t)}(1 - y_{j,T})$ is the number of patients who have completed their toxicity assessment without experiencing a DLT. The approximation in the likelihood formula is derived from a Taylor expansion: $(1-w_{j,T}p_d)^{1-\delta_{j,T}} \approx (1-p_d)^{w_{j,T}(1-\delta_{j,T})}$ at $p_d = 0$. Similarly, for the efficacy outcome, we have 
$$\begin{aligned}
L(q_d \mid \Theta_{d,E}) & \propto \prod_{j \in I_d} q_d^{\delta_{j,E}y_{j,E}}(1-q_d)^{\delta_{j,E}(1-y_{j,E})}(1-w_{j,E}p_d)^{1-\delta_{j,E}} \\
& \approx \prod_{j \in I_d} q_d^{\delta_{j,E}y_{j,E}}(1-q_d)^{\delta_{j,E}(1-y_{j,E})}(1-q_d)^{w_{j,E}(1-\delta_{j,E})}\\ 
& \approx q_d^{n_{d,E}^{(t)}}(1-q_d)^{\tilde{m}_{d,E}^{(t)}},
\end{aligned}$$
where $n_{d,E}^{(t)} = \sum_{j\in I_d} \delta_{j,E}^{(t)}y_{j,E}$ is the number of patients who have experienced responses in $I_d^{(t)}$, $\tilde{m}_{d,E}^{(t)} = m_{d,E}^{(t)} +  \sum_{j\in I_d} (1-\delta_{j,E}^{(t)})w_{j,E}^{(t)}$ is the effective number of patients who have not experienced a response, and $m_{d,E}^{(t)} = \sum_{j\in I_d} \delta_{j,E}^{(t)}(1 - y_{j,E})$ is the number of patients who have completed their efficacy assessment without a response. Based on the likelihood functions of the toxicity and efficacy probabilities with pending outcomes, the observed toxicity probability $\hat{p}_d$ and observed efficacy probability $\hat{q}_d$ can be estimated as $\tilde{p}_d^{(t)} = n_{d,T}^{(t)}/(n_{d,T}^{(t)} + \tilde{m}_{d,T}^{(t)})$ and $\tilde{q}_d^{(t)} = n_{d,E}^{(t)}/(n_{d,E}^{(t)} + \tilde{m}_{d,E}^{(t)})$ at time $t$. 

Let $c = 1$ denote the initial cohort, where each cohort represents a group of patients joining the trial over a certain period, and $d = 1$ be the pre-specified initial dose level. The TITE-STEIN design applies the same hypotheses and design parameters $\{\phi_L, \phi_U, \psi\}$ as STEIN. TITE-STEIN's dose-finding algorithm is as follows: 
\vspace{-0.05in}
\begin{itemize}
    \item[1. ] Treat the cohort $c$ at dose level $d$. 
    \item[2. ] Calculate $\tilde{p}_d^{(t)}$ and $\tilde{q}_d^{(t)}$ for the current dose level $d$ at time $t$. Make dosing decisions based on the following criteria:
    \begin{itemize}
        \item[(a)] If $\tilde{p}_d^{(t)} \geq \phi_U$, de-escalate to dose $d - 1$.
        \item[(b)] If $\tilde{p}_d^{(t)}< \phi_L$ and $\tilde{q}_d^{(t)} \geq \psi$, stay at the current dose $d$.
        \item[(c)] If $\tilde{p}_d^{(t)} \leq \phi_L$ and $\tilde{q}_d^{(t)} < \psi$, define the admissible set $A_d = \{d-1, d, d+1\}$; or if $\phi_L < \tilde{p}_d^{(t)} < \phi_U$ and $\tilde{q}_d^{(t)}< \psi$, define the admissible set $A_d = \{d-1, d\}$. Then select the dose $d^{Next}$ for the next cohort as: $$d^{Next} = \arg\max_{d' \in A_d} \Pr(q_{d'} > \psi\mid n_{d'}, n_{d', E}^{(t)}, \tilde{m}^{(t)}_{d,E}).$$ 
    \end{itemize}
    \item[3. ] Change $c$ to $c+1$ and update $d$ based on the decision in Step 2. 
    \item[4. ] Repeat Steps 1 - 3 until the maximum sample size is reached. 
\end{itemize}

To avoid assigning patients to overly toxic or futile doses, we incorporate two dose elimination criteria shown below. In the equations, $(\pi_T, \pi_E, c_T, c_E)$ are pre-specified design parameters.  In addition, we propose an accrual suspension criterion to avoid making inaccurate decisions when a large proportion of toxicity or efficacy outcomes are still pending.
\newlength{\mylength}
\settowidth{\mylength}{\textbf{(Accural)}}
\begin{itemize}[left=0em]
    \item[] \makebox[\mylength][l]{\textbf{(Safety)}} \parbox[t]{0.9\linewidth}{If $\Pr(p_d > \pi_T\mid n_{d'}, n_{d', T}^{(t)}, \tilde{m}^{(t)}_{d,T}) > c_T$, eliminate the current dose $d$ and all higher doses, i.e., $\{d, d+1, \ldots, D\}$, from the dose list;}
    \item[] \makebox[\mylength][l]{\textbf{(Futility)}} \parbox[t]{0.9\linewidth}{If $\Pr(q_d < \pi_E\mid n_{d'}, n_{d', E}^{(t)}, \tilde{m}^{(t)}_{d,E}) > c_E$, eliminate the current dose $d$ from the dose list;}
    {\textbf{(Accrual)}} \parbox[t]{0.9\linewidth}{If more than 50\% of the patients have pending toxicity or efficacy outcomes at the current dose, suspend accrual to allow more data to become available.}
\end{itemize}
The trial should be terminated early if no suitable dose is available for the next patient cohort. 

\subsection{OBD Selection with Verification}

If the maximum sample size is reached and not all doses have been eliminated by the end of the trial, we will select the optimal dose from the remaining dose levels after thoroughly assessing the toxicity and efficacy outcomes at each dose level. The initial OBD selection procedure in TITE-STEIN follows a similar approach to the one described in STEIN\cite{lin2017stein}. We perform isotonic regression on the observed toxicity probabilities $\{\hat{p}_d\}_{d=1}^D$ to obtain the isotonically transformed values $\{\tilde{p}_d\}_{d=1}^D$. For efficacy outcomes, TITE-STEIN performs $D$ unimodal isotonic regressions on the observed efficacy probabilities $\{\hat{q}_d\}_{d=1}^D$, examining each possible mode in the dose-efficacy curve. In the $d'$-th model, dose level $d'$ attains the highest efficacy probability, $d' = 1, \ldots, D$. In other words, after unimodal isotonic transformations, the transformed efficacy probabilities $\{\tilde{q}_{d'd}\}_{d=1}^D$ in the $d'$-th model satisfy $\tilde{q}_{d'1} \leq \cdots \leq \tilde{q}_{d'd'}$ and $\tilde{q}_{d'd'} \geq \cdots \geq \tilde{q}_{d'D}$. The pseudo-likelihood based on the $d'$-th unimodal isotonic regression is given by 
$$
\tilde{L}_{d'} = \prod_{d = 1}^D \binom{n_d}{y_{d, E}}\tilde{q}_{d'd}^{y_{d, E}}(1-\tilde{q}_{d'd}^{y_{d, E}})^{n_j - y_{d,E}}.
$$
The final model-averaged estimate of $q_d$ is computed as $\tilde{q}_d = \sum_{d' = 1}^D \pi_{d'}\tilde{q}_{d'd}$, where $\pi_{d'} = \tilde{L}_{d'}/\sum_{d=1}^D \tilde{L}_d$. Following Lin and Yin \cite{lin2017stein}, the optimal dose $d^*$ is the dose with the highest utility and is given by
$$
d^* = \max_{d} \{U(\tilde{p}_d, \tilde{q}_d)\}= \max_{d}\{\tilde{q}_d - w_1\tilde{p}_d - w_2\tilde{p}_dI(\tilde{p}_d > \phi)\},
$$
where the two design parameters $w_1 = 0.33$ and $w_2 = 1.09$ are as recommended in the original STEIN paper \cite{lin2017stein}.

Although the two dose elimination criteria used during the dose exploration phase help exclude overly toxic or inefficacious dose levels before the initial OBD selection, they alone cannot guarantee that the selected optimal dose $d^*$ is truly admissible. Even if $d^*$ has the highest utility value among the remaining doses, its utility value, $U(p_{d^*}, q_{d^*})$, may still be low. This may indicate that the dose has a poor clinical utility and suggest that it should not be chosen as the OBD. To address this issue, following the PRINTE design\citet{lin2021probability}, we propose an additional OBD verification procedure after the initial OBD selection to confirm whether the selected dose $d^*$ is appropriate as the OBD: 
\begin{itemize}
    \item[(1)] Generate a total of $M$ independent random samples from the posterior distributions of toxicity and efficacy probabilities for all dose levels. Specifically, $p_{d}^{(m)} \sim \text{Beta}(n_{d,T} + \alpha_{T}, n_{d} - n_{d,T} + \beta_T)$ and $q_{d}^{(m)} \sim \text{Beta}(n_{d,E} + \alpha_{E}, n_{d} - n_{d,E} + \beta_E)$ for $m = 1, \ldots, M$. 
    \item[(2)] For each sample $m$, perform an isotonic regression on $\{p_d^{(m)}\}_{d=1}^D$ to obtain the isotonically transformed values $\{\tilde{p}_d^{(m)}\}_{d=1}^D$. For the efficacy probabilities $\{q_d^{(m)}\}_{d=1}^D$, apply $D$ unimodal isotonic regressions to derive the model-averaged efficacy probability estimates $\{\tilde{q}_{d}^{(m)}\}_{d=1}^D$, where $\tilde{q}_{d}^{(m)} = \sum_{d'=1}^D \pi_{d'}^{(m)}\tilde{q}_{d'd}^{(m)}$.
    \item[(3)] For the selected dose level $d^*$, obtain a total of $M$ samples $\{U(\tilde{p}_{d^*}^{(m)}, \tilde{q}_{d^*}^{(m)})\}_{m=1}^M$ where $$U(\tilde{p}_{d^*}^{(m)}, \tilde{q}_{d^*}^{(m)}) = \tilde{q}_{d^*}^{(m)} - w_1\tilde{p}_{d^*}^{(m)} - w_2\tilde{p}_{d^*}^{(m)}I(\tilde{p}_{d^*}^{(m)} > \phi).$$ 
    \item[(4)] Calculate the p-value of $U(p_{d^*}, q_{d^*})$ being greater than a pre-specified utility cutoff value $U_B$ given by
    $$
    p_{g} = \Pr(U(p_{d^*}, q_{d^*}) > U_B) = M^{-1}\sum_{m=1}^M I(U(p_{d^*}^{(m)}, q_{d^*}^{(m)}) > U_B).
    $$
    If $p_g > p_{min}$ where $p_{min}$ is also a pre-specified design parameter, select $d^*$ as the OBD; otherwise, do not select any dose as the OBD at the end of the trial. 
\end{itemize}

Selecting an overly toxic or inefficacious dose with a low clinical utility as the OBD can result in the failure of later-phase trials, waste valuable resources, and compromise patient safety. Our OBD verification procedure reduces the risk of selecting an unsuitable dose as the OBD, as demonstrated in the following simulation section.

\section{Simulation Studies}\label{sec3:simu}

\subsection{Simulation Settings}
In our simulation study, we considered $D = 5$ dose levels, with the assumed dosages of 100mg, 200mg, 300mg, 400mg, and 500mg. The assumption of certain dosage values is needed for the Joint TITE-CRM and LO-TC designs. The maximum number of cohorts was set to $C_M = 15$ with a cohort size of $n_c = 3$, resulting in a maximum sample size of 45. The accrual rate was set to three patients per month. We assumed the maximum acceptable toxicity probability to be $p_T = 0.3$, and the minimum acceptable efficacy probability to be $q_E = 0.25$. The assessment windows for toxicity and efficacy set to 1 month ($A_T = 1$) and 3 months ($A_E = 3$), respectively. We assumed a 1-month treatment cycle, an assumption required only for the Joint TITE-CRM design. The time to toxicity and efficacy outcomes were generated using two uniform distributions: $X_{j,T} \sim U(0, A_T)$ and $X_{j,E} \sim U(0, A_E)$.  

\begin{figure}
    \centering
    \includegraphics[width = 7in, height = 4.6in]{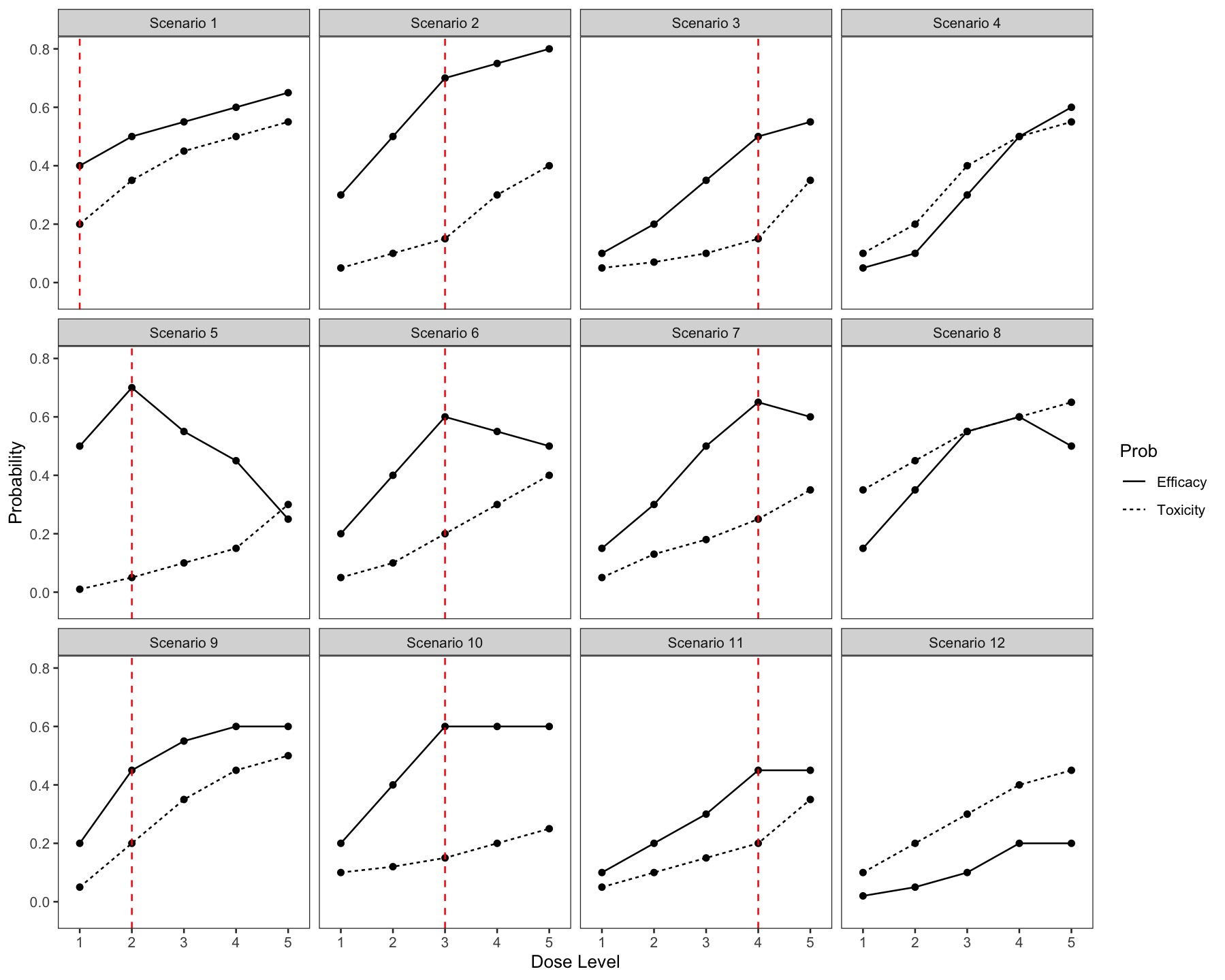}
    \caption{True toxicity and efficacy probability curves with the OBD (vertical red line) for all simulation scenarios}
    \label{fig:true_prob}
    \begin{minipage}{\textwidth} 
    \footnotesize
    \vspace{0.2in}
\end{minipage}
\end{figure}

We compared TITE-STEIN with three model-assisted designs, STEIN, TITE-BOIN12, and TITE-BOIN-ET, and two model-based designs, LO-TC and Joint TITE-CRM. Except for STEIN, all these designs utilized late-onset toxicity and efficacy outcomes to identify the OBD. We evaluated twelve different scenarios representing three types of dose-response relationships: (1) Increasing (scenarios 1, 2, 3, and 4), (2) Unimodal (scenarios 5, 6, 7, and 8), and (3) Plateau (scenarios 9, 10, 11, and 12). We assumed that the OBD did not exist in scenarios 4, 8, and 12. Each simulation was independently replicated 1000 times. The true probability curves and values of toxicity and efficacy for all scenarios are shown in Figure~\ref{fig:true_prob} and Table~\ref{tab:true_simuvalue}, respectively. 

For the key design parameters of STEIN and TITE-STEIN, we set $\phi = 0.3$, $\phi_1 = 0.75\phi = 0.225$, $\phi_2 = 1.25\phi = 0.375$, $\psi_1 = 0.3$, $\psi_2 = 0.8$, $w_1 = 0.33$, $w_2 = 1.09$, following the recommended values in \citet{lin2017stein}. The same $(w_1, w_2) = (0.33, 1.09)$ values were also utilized in the utility function of Joint TITE-CRM. For the OBD verification procedure of TITE-STEIN, we set the random sample size $M = 1000$, utility cutoff value $U_B = \psi_1 - w_1\phi = 0.201$, and $p_{min} = 0.1$. In TITE-BOIN12, we assumed the utility scores $\{u_{11}, u_{10}, u_{01}, u_{00}\} = \{60, 0, 100, 40\}$, as recommended by \citet{zhou2022tite}. With $p_T = 0.3$, the escalation and de-escalation boundaries of TITE-BOIN12 were set to $\lambda_e = 0.236$ and $\lambda_d = 0.359$, respectively. For TITE-BOIN-ET, we assumed the target toxicity probability to be $\phi^{ET} = 0.3$, the target efficacy probability to be $\delta^{ET} = 0.6$, $\phi_1^{ET} = 0.1\phi$, $\phi_2^{ET} = 1.4\phi$, and $\delta_1^{ET} = 0.6\delta^{ET}$, so that the optimal design parameter values $\{\lambda_1^{ET}, \lambda_2^{ET}, \eta_1^{ET}\}$ were $\{0.14, 0.35, 0.47\}$. The prior distribution parameters used in both designs were consistent with those used in \citet{jin2014using} and  \citet{barnett2024joint}. For LO-TC, we used pairs $(\pi_{Eff}, \pi_{Tox}) = (0.3, 0), (0.56, 0.3), (0.8, 1)$ to define the trade-off contour. For Joint TITE-CRM, we set a follow-up period of $\tau = 3$ cycles, with each cycle lasting one month. The target toxicity probability for Joint TITE-CRM was $\pi_T^* = 0.391$, corresponding to $p_T = 0.3$ in the first cycle. These aforementioned design parameters specified for each unique design were consistent with those in the original papers. Altering them could change the dosing decisions. The two dose elimination criteria mentioned earlier are adopted for all these TITE designs with the same design parameters for consistency. Specifically, we set $(\pi_T, \pi_E, c_T, c_E) = (0.3, 0.25, 0.95, 0.9)$ for the safety and futility criteria introduced in Section~\ref{TITE-STEIN}. We also applied a consistent accrual suspension criterion 
for all the TITE designs except for Joint TITE-CRM, where it would significantly alter dose exploration. 

\subsection{Simulation Results}\label{sec_simu_results}
The simulation results are presented in Table~\ref{tab:simuresult}, which includes the percentage of times each dose level was selected as the OBD, the probability of terminating a trial early without selecting any dose as the OBD, the number of patients assigned to each dose level, and the overall trial duration. When the OBD existed, TITE-STEIN and STEIN demonstrated comparable performance in OBD selection probability and patient allocation. However, TITE-STEIN had a shorter trial duration by utilizing late-onset outcomes compared to STEIN. When the OBD did not exist, TITE-STEIN outperformed STEIN in accurately terminating the trial without selecting a dose. Furthermore, compared to other late-onset designs, TITE-STEIN selected the OBD more accurately and allocated more patients to the OBD in most scenarios. Four late-onset designs, TITE-BOIN12, TITE-BOIN-ET, TITE-STEIN, and LO-TC, which adopted the same accrual suspension criterion, had similar trial durations across all scenarios. Joint TITE-CRM, making the dosing decision after each patient cohort's first cycle and in the absence of the accrual suspension rule, resulted in the shortest trial duration.


Scenarios 1 to 4 considered an increasing dose-response relationship. Scenario 1, where dose level 1 was the OBD and the MTD, showed that TITE-STEIN exhibited the highest probability of accurately selecting the OBD and allocated the largest number of patients to it. Other designs had a probability exceeding 30\% of selecting a toxic dose as the OBD, while TITE-BOIN-ET and Joint TITE-CRM showed an early termination probability of over 12\% without selecting any dose as the OBD. In scenario 2, where $d_{OBD} = 3$ and $d_{MTD} = 4$, TITE-STEIN outperformed other late-onset designs in selecting the OBD accurately. TITE-BOIN-ET and TITE-STEIN allocated more patients to the OBD than the other late-onset designs. LO-TC had a 37.7\% probability of selecting dose level 5, a toxic dose, as the OBD, with an average of 13.3 patients assigned to that dose. In scenario 3, dose level 4 was both the $d_{OBD}$ and the $d_{MTD}$. TITE-STEIN still had the highest probability of accurately identifying dose level 4 as the OBD and assigned most patients to that dose. LO-TC had an extremely high probability of selecting dose level 5, with an average of nearly 30 patients assigned to this toxic dose. Scenario 4 did not have any OBD and dose level 2 was the MTD. Among the six designs considered, TITE-STEIN had a high probability of terminating the trial without selecting any dose as the OBD with an effective overdose control, only slightly lower than that of Joint TITE-CRM. In this scenario, STEIN had a 41.4\% probability of terminating the trial without selecting a dose as the OBD, which was 15.2\% lower than that of TITE-STEIN.

Scenarios 5 to 8 considered a unimodal dose-response relationship. In scenario 5, $d_{OBD} = 2$ and $d_{MTD} = 5$. All the model-assisted designs had a probability greater than 64\% of correctly selecting the OBD, among which TITE-STEIN had the best performance. TITE-STEIN also assigned the most patients to the OBD among the late-onset designs. However, LO-TC had a 51.2\% probability of selecting dose level 3 as the OBD, while Joint TITE-CRM had over an 80\% probability of selecting dose level 5 as the OBD. In scenario 6 where $d_{OBD} = 3$ and $d_{MTD} = 4$, TITE-BOIN-ET had a slightly higher probability of selecting the OBD compared to STEIN and TITE-STEIN, but it assigned fewer patients to the OBD. In scenario 7, dose level 4 was both the OBD and the MTD. Although TITE-STEIN had a lower OBD selection probability compared to TITE-BOIN12, TITE-BOIN-ET, and Joint TITE-CRM, it still had the most effective overdose control. TITE-BOIN-ET had an 11.4\% probability of selecting a toxic dose, despite having the highest OBD selection probability. In scenario 8, with no OBD or MTD, TITE-BOIN-ET, TITE-STEIN, and Joint TITE-CRM each had over an 80\% probability of not selecting any dose as the OBD, while LO-TC had the lowest probability. 

Scenarios 9 to 12 considered a plateau dose-response relationship. In scenario 9, dose level 2 was both the OBD and the MTD. TITE-STEIN had the highest OBD selection probability and the lowest probability of selecting a toxic dose as the OBD, while Joint TITE-CRM assigned slightly more patients to the OBD. In scenario 10, with $d_{OBD} = 2$ and $d_{MTD} = 5$, TITE-STEIN demonstrated higher OBD selection accuracy and allocated most patients to the OBD among the late-onset designs. Both LO-TC and Joint TITE-CRM had over a 69\% probability of selecting dose level 5 as the OBD. In scenario 11, with dose level 4 as both the OBD and MTD, TITE-STEIN outperformed other late-onset designs in accurate OBD selection and assigned most patients to it. Finally, in scenario 12, where the OBD did not exist and the MTD was dose level 3, TITE-STEIN, LO-TC, and Joint TITE-CRM each had over a 70\% probability of correctly not selecting any dose as the OBD. In contrast, TITE-BOIN-ET had only a 34.5\% probability of not selecting any dose as the OBD.

\subsection{Sensitivity Analysis}
We performed three independent sensitivity analyses (SA) to evaluate the robustness of our simulation findings by modifying the following design parameters respectively: 
\begin{itemize}
    \item[(SA1)] the maximum number of cohorts was set to $C_M = 30$; 
    \item[(SA2)] the toxicity assessment window was set to $A_T = 2$ months and the efficacy assessment window was set to $A_E = 6$ months; 
    \item[(SA3)] the starting dose level to which the initial cohort of patients was assigned to dose level 2. 
\end{itemize}

By increasing the maximum sample size in SA1, all the model-assisted designs evaluated identified the OBD more accurately when it existed and terminated the trial without selecting any dose when the OBD did not exist in most scenarios. However, LO-TC and Joint TITE-CRM tended to choose a higher dose level as the OBD in certain scenarios, such as in scenario 10. SA2 had extended toxicity and efficacy assessment windows， which led to longer trial durations across all designs except Joint TITE-CRM. This exception occurred for Joint TITE-CRM because it made dosing decisions after the first cycle for each cohort \cite{barnett2024joint}, without accounting for the duration of the assessment window of either outcome. Since the assessment windows were longer in this case, a smaller proportion of information was observed within one cycle. This led Joint TITE-CRM to select the OBD less accurately and to allocate fewer patients to the OBD. In contrast, results of the OBD selection probability and patient allocation to the OBD remained consistent with those of the original simulation for the other five designs, suggesting that TITE-STEIN is robust to variations in the duration of the assessment windows of the outcomes. In SA3, the starting dose level was set to dose level 2. With $d_{OBD} = 1$ in scenario 1, the OBD selection probabilities of all model-assisted designs were comparable to those of the original simulations. However, for LO-TC and Joint TITE-CRM, the OBD selection probabilities and patient allocation to the OBD both decreased. In the other scenarios where the OBD was above dose level 1, these designs showed a slight improvement by selecting the OBD more accurately and allocating more patients to the OBD compared to the original simulation results. In summary, the sensitivity analyses confirmed the reliability and robustness of our simulation findings. Detailed results from the sensitivity analyses are provided in Section 2 of the supplementary document. 


\section{Case Study}\label{sec4:case}

We conducted a case study using data from the TRANSCEND NHL 001 trial of lisocabtagene maraleucel (liso-cel) for patients with relapsed or refractory B-cell lymphomas  \cite{abramson2020lisocabtagene}. Liso-cel is an autologous, chimeric antigen receptor (CAR) T-cell therapy targeting CD19. In this study, patients were assigned to one of three target dose levels: $50 \times 10^6$ CAR$^+$ T cells (DL1), $100 \times 10^6$ CAR$^+$ T cells (DL2), and $150 \times 10^6$ CAR$^+$ T cells (DL3). From Jan 11, 2016, to July 5, 2019, 269 patients received at least one infusion of liso-cel. DLT was evaluated in the trial from the time of liso-cel infusion through 28 days, indicating that the toxicity assessment window is $A_T = 28$ days. In our analysis, we used the DLT rate and the objective response rate (ORR) as the toxicity and efficacy outcomes. The objective response, assessed per Lugano criteria \cite{cheson2014recommendations}, was defined as either a complete response or a partial response after the infusion. The evaluation of objective response was from liso-cel infusion to disease progression, the end of the study, or the start of another anticancer therapy, or hematopoietic stem-cell transplant (HSCT), whichever event occurred first. In liso-cel, as observed in other CAR T-cell therapies, patients typically responded within the first few months, with the vast majority of responders occurring within the first month. Considering this, we assumed the efficacy assessment window to be $A_E = 84$ days, equivalent to three times the toxicity assessment window. We also assumed the probabilities of the response occurring in [0, 28] days, [29, 56] days, and [57, 84] days to be 0.7, 0.2, and 0.1, respectively. Within each time interval, the time to the outcome was assumed to follow a uniform distribution. 

This trial consisted of three stages: dose finding, dose expansion, and dose confirmation. During the dose finding and dose expansion stages, all three dose levels were assessed, but only DL2 proceeded to the dose confirmation stage as the optimal dose. 54 patients in the dose-finding stage and 83 patients in the dose-expansion stage were enrolled. The TRANSCEND NHL-001 trial used the mCRM design for dose finding \cite{quintana2016bayesian}. Given these trial data, we made some reasonable assumptions to redesign this trial, and applied TITE-STEIN. We assumed an early-phase trial with the maximum number of cohorts to be 45 and a cohort size of 3, resulting in a maximum sample size of 135, to approximate the sample size of 137 across the two stages. We assumed an accrual rate of 1 patient per 5 days. We combined the toxicity and efficacy data across all three stages and summarized them in Table~\ref{tab:case_summary}. Based on the observed data in Table~\ref{tab:case_summary}, we assumed the DLT rates in DL1, DL2, and DL3 to be 7\%, 10\%, and 12\%, and the ORR in DL1, DL2, and DL3 to be 65\%, 75\%, and 75\%, and thus assumed a plateau dose-response relationship. We set the maximum acceptable toxicity probability $p_T = 0.3$ and the minimum acceptable efficacy probability $q_E = 0.25$. DL2 was identified as the OBD by TITE-STEIN based on the assumed toxicity and efficacy probability values. This is consistent with the decision to select DL2 for the dose confirmation stage in the TRANSCEND NHL 001 trial.

We compared the performance of three model-assisted TITE designs, TITE-BOIN12, TITE-BOIN-ET, and TITE-STEIN. Their results are summarized in Table~\ref{tab:case_result}. TITE-STEIN had the highest probability of selecting the OBD and assigned most patients to the OBD. However, TITE-BOIN-ET tended to select DL1 and assigned more patients to DL1 compared to the other two designs. The three designs exhibited similar trial durations. In the TRANSCEND NHL-001 trial, enrollment was gated in the dose-finding portion of the trial, and there was no assumption of cohort size. However, these three considered TITE designs assume a cohort size and require dosing decisions to be made for each new patient cohort. Due to the implementation of the accrual suspension rule in these designs, new patients had to wait until sufficient data from pending outcomes became available, particularly when the number of enrolled patients was low. As a result, the enrollment duration for these TITE designs is generally much longer than the trial duration of the TRANSCEND NHL-001 trial. This case study demonstrates that when applied to this CAR T-cell therapy trial,  TITE-STEIN more accurately identifies the optimal dose and allocates patients more effectively to the true OBD compared with TITE-BOIN12 and TITE-BOIN-ET.


\section{Software}\label{sec5:software}
The software for implementing TITE-BOIN12 is available on \href{https://trialdesign.org/}{www.trialdesign.org}. The TITE-BOIN-ET program is provided at \href{https://github.com/yamagubed/boinet}{https://github.com/yamagubed/boinet}, along with an R package called ``boinet''. The Joint TITE-CRM program is available at \href{https://github.com/helenyb/DF_Late_Activity_Safety}{https://github.com/helenyb/DF\_Late\_Activity\_Safety}. The R programs of TITE-STEIN and other designs used in our simulation studies and sensitivity analyses are shared at \href{https://github.com/EugeneHao/TITE-STEIN}{https://github.com/EugeneHao/TITE-STEIN}.

\section{Discussion}\label{sec6:discussion}






We have proposed TITE-STEIN, a novel model-assisted dose-finding design, to address the challenges of making timely dosing decisions in trials with fast patient accrual or late-onset toxicity and efficacy outcomes. The development of TITE-STEIN is motivated by the favorable statistical operating characteristics of STEIN, as demonstrated by \citet{sun2024statistical}. Their work highlights that STEIN consistently identifies the OBD accurately across various scenarios, while simultaneously minimizing patient exposure to overly toxic doses. Furthermore, TITE-STEIN incorporates a verification step before the final OBD selection, mitigating the risk of choosing an inadmissible dose when the true OBD does not exist. Compared to STEIN, TITE-STEIN significantly increases trial efficiency by shortening the overall trial duration without compromising the accuracy of OBD selection. Moreover, when the OBD does not exist, TITE-STEIN is more likely to terminate the trial without selecting a suboptimal dose compared to STEIN.

We evaluated the statistical operating characteristics of TITE-STEIN, comparing its performance to other OBD-finding TITE designs, namely TITE-BOIN12, TITE-BOIN-ET, LO-TC, and Joint TITE-CRM, through extensive simulation studies and sensitivity analyses. The results demonstrated that TITE-STEIN outperformed these designs in most scenarios. Specifically, when the true OBD existed, TITE-STEIN more accurately selected the OBD and allocated more patients to that OBD compared to the other designs. Furthermore, in the absence of the true OBD, TITE-STEIN had a higher probability of terminating the trial early without selecting any dose, compared to the other designs. Additionally, TITE-STEIN consistently exhibited the best overdose control among the designs evaluated by allocating the smallest number of patients to toxic doses. Its trial durations were comparable to those of TITE-BOIN12, TITE-BOIN-ET, and LO-TC, while Joint TITE-CRM had the shortest durations since an accrual suspension rule was not implemented for this design. In the case study using TRANSCEND NHL 001 trial data, TITE-STEIN selected the OBD more accurately and allocated more patients to the OBD compared to TITE-BOIN12 and TITE-BOIN-ET.

Several directions for future research are apparent. First, the current TITE-STEIN implementation may encounter challenges in escalating from initial dose level(s), particularly in situations where all the observed efficacy probabilities are high (e.g., $>50$\%) or when the dose-efficacy curve exhibits multiple peaks \cite{shi2021utpi}. Future work could explore extensions to accommodate these more complex dose-efficacy patterns. Second, the current design assumes independence between toxicity and efficacy outcomes, which may not hold in some therapeutic contexts, such as those involving immune checkpoint inhibitors\cite{rong2024association}. Jointly modeling correlated toxicity and efficacy outcomes in these scenarios could be considered. For example, the U-BOIN design, proposed by \citet{zhou2019utility}, utilizes a multinomial-Dirichlet model to jointly model toxicity and efficacy. Third, incorporating historical data, as suggested by \citet{zhou2021incorporating}, into TITE-STEIN may further improve its efficiency and adaptability. For example, the investigational drug may have been studied previously in other indications, or similar agents from the same drug class may have been evaluated in earlier phase I trials \cite{zohar2011approach}. Fourth, the current TITE-STEIN design considers only binary outcomes for toxicity and efficacy. Extending TITE-STEIN to utilize continuous or ordinal toxicity and/or efficacy endpoints could reduce information loss and enhance the accuracy of OBD selection. Several methods, such as total toxicity burden \citep{bekele2004dose}, equivalent toxicity score \citep{chen2010novel}, and total toxicity profile \citep{ezzalfani2013dose}, have been proposed to handle continuous or ordinal outcomes. For example, TITE-gBOIN-ET \citep{takeda2023tite} uses normalized equivalent toxicity and efficacy scores as quasi-Bernoulli endpoints. Finally, integrating pharmacokinetic (PK) outcomes into TITE-STEIN could further enhance trial efficiency and performance. For instance, TITE-PKBOIN-12 \cite{sun2023pkboin} has demonstrated improved accuracy in OBD selection and patient allocation to the OBD compared to TITE-BOIN12 by incorporating PK information. Similar extensions to TITE-STEIN could be explored for trial settings where PK data are available and relevant.
\bibliographystyle{agsm}
\bibliography{cite}

\newpage

\begin{table}
\centering
\caption{TITE-STEIN  decision table with $\{\phi, \phi_1, \phi_2, \psi_1, \psi_2\} = \{0.3, 0.225, 0.375, 0.3, 0.8\}$ }\label{tab:TITE_STEIN_dec}
\begin{threeparttable}
\begin{tabular}{||c|rr|rr|c||c|rr|rr|c|}
\hline
$\bm{n_d}$ & $\bm{n_{d,T}^{(t)}}$ & $\bm{\tilde{m}_{d,T}^{(t)}}$ & $\bm{n_{d,E}^{(t)}}$ & $\bm{\tilde{m}_{d,E}^{(t)}}$ & \textbf{Decision} & $\bm{n_d}$ & $\bm{n_{d,T}^{(t)}}$ & $\bm{\tilde{m}_{d,T}^{(t)}}$ & $\bm{n_{d,E}^{(t)}}$ & $\bm{\tilde{m}_{d,E}^{(t)}}$ & \textbf{Decision} \\ 
\hline
3 & \multicolumn{5}{c||}{Pending if $\max\{o_{d,T}^{(t)}, o_{d,E}^{(t)}\} \geq 2$} & 9 & $3$ & $ \leq 5.90 $ & \multicolumn{2}{c|}{Any} & D   \\
\hline 
3 & $\geq 2$ & $\leq 0.46$ & \multicolumn{2}{c|}{Any} & DU & 9 & $3$ & $ > 5.90 $ & $\geq 5$ & $< 3.91$ & S \\
\hline
3 & $1$ & $\leq 1.96$ & \multicolumn{2}{c|}{Any} & D & 9 & $3$ & $ > 5.90 $ & 5 & $\geq 3.91$ & TBD \\
\hline
3 & $\leq 1$ & $> 1.96$ & $\geq 2$ & Any & S & 9 & $3$ & $ > 5.90 $ & 4 & $< 3.13$ & S\\
\hline
3 & $\leq 1$ & $> 1.96$ & $\leq 1$ & Any & TBD & 9 & $3$ & $ > 5.90 $ & 4 & $\geq 3.13$ & TBD \\
\hline
6 & \multicolumn{5}{c||}{Pending if $\max\{o_{d,T}^{(t)}, o_{d,E}^{(t)}\} \geq 4$} & 9 & $3$ & $ > 5.90 $ & 3 & $< 2.34$ & S \\
\hline
6 & $\geq 4$ & $\leq 1.86$ & \multicolumn{2}{c|}{Any} & DU & 9 & $3$ & $ > 5.90 $ & 3 & $\geq 2.34$ & TBD \\
\hline
6 & 3 & $\leq 1.53$ & \multicolumn{2}{c|}{Any} & DU & 9 & $3$ & $ > 5.90 $ & $\leq 2$ & Any & TBD \\
\hline
6 & 3 & $> 1.53$ & \multicolumn{2}{c|}{Any} & D & 9 & 2 & $ \leq 3.93 $ & \multicolumn{2}{c|}{Any} & D \\
\hline
6 & 2 & $\leq 3.93$ & \multicolumn{2}{c|}{Any} & D & 9 & 2 & $ > 3.93 $ & $\geq 5$ & $< 3.91$ & S \\
\hline 
6 & 2 & $> 3.93$ & $\geq 3$ & $<2.34$ & S & 9 & 2 & $ > 3.93 $ & 5 & $\geq 3.91$ & TBD \\
\hline
6 & 2 & $> 3.93$ & $3$ & $\geq 2.34$ & TBD & 9 & 2 & $ > 3.93 $ & 4 & $< 3.13$ & S \\
\hline 
6 & 2 & $> 3.93$ & $2$ & $< 1.56$ & S & 9 & 2 & $ > 3.93 $ & 4 & $\geq 3.13$ & TBD\\
\hline
6 & 2 & $> 3.93$ & $2$ & $\geq 1.56$ & TBD & 9 & 2 & $ > 3.93 $ & 3 & $< 2.34$ & S  \\
\hline
6 & 2 & $> 3.93$ & $\leq 1$ & Any & TBD & 9 & 2 & $ > 3.93 $ & 3 & $\geq 2.34$ & TBD \\
\hline 
6 & $\leq 1$ & Any & $\geq 3$ & $< 2.34$ & S & 9 & 2 & $ > 3.93 $ & $\leq 2$ & Any & TBD \\
\hline
6 & $\leq 1$ & Any & $3$ & $\geq 2.34$ & TBD & 9 & $\leq 1$ & Any & $\geq 5$ & $< 3.91$ & S \\
\hline 
6 & $\leq 1$ & Any & $2$ & $< 1.56$ & S & 9 & $\leq 1$ & Any & 5 & $\geq 3.91$ & TBD \\
\hline
6 & $\leq 1$ & Any & $2$ & $\geq 1.56$ & TBD & 9 & $\leq 1$ & Any & 4 & $< 3.13$ & S \\
\hline
6 & $\leq 1$ & Any & $\leq 1$ & Any & TBD & 9 & $\leq 1$ & Any & 4 & $\geq 3.13$ & TBD \\
\hline
9 & \multicolumn{5}{c||}{Pending if $\max\{o_{d,T}^{(t)}, o_{d,E}^{(t)}\} \geq 5 $} & 9 & $\leq 1$ & Any & 3 & $< 2.34$ & S \\
\hline
9 & $\geq 5$ & Any & \multicolumn{2}{c|}{Any} & DU & 9 & $\leq 1$ & Any & 3 & $\geq 2.34$ & TBD \\
\hline
9 & $4$ & $\leq 2.76$ & \multicolumn{2}{c|}{Any} & DU & 9 & $\leq 1$ & Any & $\leq 2$ & Any & TBD \\
\hline
9 & $4$ & $> 2.76$ & \multicolumn{2}{c|}{Any} & D & \multicolumn{6}{c|}{} \\
\hline
\end{tabular}
\begin{tablenotes}
\footnotesize
\setlength{\baselineskip}{0.6\baselineskip}
\item[(a)] $n_d$: number of patients; $n_{d,T}^{(t)}$: number of patients with DLTs; $\tilde{m}_{d,T}^{(t)}$: effective number of patients who have not experienced DLT; $n_{d,E}^{(t)}$: number of patients with response; $\tilde{m}_{d,T}^{(t)}$: effective number of patients who have not experienced response;
\item[(b)] D = de-escalate to the next lower dose; DU = de-escalate to the next lower dose and terminate the current and higher doses; S = stay at the current dose; TBD = to be determined based on the posterior probabilities of the doses within the admissible set.
\end{tablenotes}
\end{threeparttable}
\end{table}

\begin{table}
\centering
\caption{True toxicity and efficacy probability values for all simulation scenarios}\label{tab:true_simuvalue}
\begin{threeparttable}
\begin{tabular}{|l|ccccc|ccccc|}
\hline
Category & \multicolumn{10}{c|}{Dose Level}\\
\hline
  & DL1 & DL2 & DL3 & DL4 & DL5 & DL1 & DL2 & DL3 & DL4 & DL5 \\
\hline
&  \multicolumn{5}{c|}{Scenario 1 ($d_{OBD}$ = 1)} & \multicolumn{5}{c|}{Scenario 2 ($d_{OBD}$ = 3)} \\
Toxicity & \textbf{0.20} & 0.35 & 0.45 & 0.50 & 0.55 & 0.05 & 0.10 & \textbf{0.15} & 0.30 & 0.40 \\
Efficacy & \textbf{0.40} & 0.50 & 0.55 & 0.60 & 0.65 & 0.30 & 0.50 & \textbf{0.70} & 0.75 & 0.80 \\
\hline
& \multicolumn{5}{c|}{Scenario 3 ($d_{OBD}$ = 4)} & \multicolumn{5}{c|}{Scenario 4 (No OBD)} \\ 
Toxicity & 0.05 & 0.07 & 0.10 & \textbf{0.15} & 0.35 & 0.10 & 0.20 & 0.40 & 0.50 & 0.55 \\
Efficacy & 0.10 & 0.20 & 0.35 & \textbf{0.50} & 0.55 & 0.05 & 0.10 & 0.30 & 0.50 & 0.60 \\
\hline
& \multicolumn{5}{c|}{Scenario 5 ($d_{OBD}$ = 2)} & \multicolumn{5}{c|}{Scenario 6 ($d_{OBD}$ = 3)} \\
Toxicity & 0.01 & \textbf{0.05} & 0.10 & 0.15 & 0.30 & 0.05 & 0.10 & \textbf{0.20} & 0.30 & 0.40 \\
Efficacy & 0.50 & \textbf{0.70} & 0.55 & 0.45 & 0.25 & 0.20 & 0.40 & \textbf{0.60} & 0.55 & 0.50 \\
\hline
& \multicolumn{5}{c|}{Scenario 7 ($d_{OBD}$ = 4)} & \multicolumn{5}{c|}{Scenario 8 (No OBD)} \\
Toxicity & 0.05 & 0.13 & 0.18 & \textbf{0.25} & 0.35 & 0.35 & 0.45 & 0.55 & 0.60 & 0.65 \\
Efficacy & 0.15 & 0.30 & 0.50 & \textbf{0.65} & 0.60 & 0.15 & 0.35 & 0.55 & 0.60 & 0.50 \\
\hline
& \multicolumn{5}{c|}{Scenario 9 ($d_{OBD}$ = 2)} & \multicolumn{5}{c|}{Scenario 10 ($d_{OBD}$ = 3)} \\
Toxicity & 0.05 & \textbf{0.20} & 0.35 & 0.45 & 0.50 & 0.10 & 0.12 & \textbf{0.15} & 0.20 & 0.25 \\
Efficacy & 0.20 & \textbf{0.45} & 0.55 & 0.60 & 0.60 & 0.20 & 0.40 & \textbf{0.60} & 0.60 & 0.60 \\
\hline
& \multicolumn{5}{c|}{Scenario 11 ($d_{OBD}$ = 4)} & \multicolumn{5}{c|}{Scenario 12 (No OBD)} \\
Toxicity & 0.05 & 0.10 & 0.15 & \textbf{0.20} & 0.35 & 0.10 & 0.20 & 0.30 & 0.40 & 0.45 \\
Efficacy & 0.10 & 0.20 & 0.30 & \textbf{0.45} & 0.45 & 0.02 & 0.05 & 0.10 & 0.20 & 0.20 \\ 
\hline
\end{tabular}
\begin{tablenotes}
\footnotesize
\setlength{\baselineskip}
{0.6\baselineskip}
\vspace{0.1in}
\item[(a)] Scenarios 1 - 4: increasing relationship;  
\item[(b)] Scenarios 5 - 8: unimodal relationship; 
\item[(c)] Scenarios 9 - 12: plateau relationship.
\end{tablenotes}
\end{threeparttable}
\end{table}

\clearpage
\setlength{\LTleft}{-1cm} 
\begin{longtable}{|l|cccccc|ccccc|c|}
\caption{Simulation results for all designs with $p_T = 0.3$, $q_E = 0.25$}\label{tab:simuresult}\\
\hline
\multirow{2}{*}{} & \multicolumn{6}{c|}{Selection Probability (\%)} & \multicolumn{5}{c|}{Number of Assigned Patients} & Duration\\
\cline{1-13}
Dose Level & DL1 & DL2 & DL3 & DL4 & DL5 & ET & DL1 & DL2 & DL3 & DL4 & DL5 & (Months)\\
\hline
Design & \multicolumn{12}{c|}{Scenario 1 ($d_{OBD} = 1$)} \\
\hline
STEIN & \textbf{68.9} & 24.0 & 3.0 & 0.3 & 0.0 & 3.8 & \textbf{24.4} & 15.4 & 3.5 & 0.4 & 0.0 & 53.5 \\ 
  TITE-BOIN12 & \textbf{63.0} & 27.1 & 5.2 & 0.4 & 0.1 & 4.2 & \textbf{24.6} & 13.5 & 4.4 & 1.1 & 0.2 & 24.8 \\ 
  TITE-BOIN-ET & \textbf{44.4} & 37.3 & 4.8 & 0.6 & 0.2 & 12.7 & \textbf{20.3} & 16.3 & 5.6 & 1.4 & 0.3 & 25.0 \\ 
  TITE-STEIN & \textbf{70.7} & 22.3 & 3.3 & 0.1 & 0.0 & 3.6 & \textbf{25.5} & 15.1 & 2.9 & 0.4 & 0.0 & 23.9 \\ 
  LO-TC & \textbf{29.0} & 38.5 & 20.0 & 6.8 & 3.8 & 1.9 & \textbf{10.4} & 13.6 & 10.3 & 5.2 & 4.8 & 26.8 \\ 
  Joint TITE-CRM & \textbf{54.3} & 27.6 & 1.7 & 0.4 & 0.0 & 16.0 & \textbf{23.4} & 13.2 & 2.3 & 0.6 & 0.1 & 15.1 \\ 
\hline
Design & \multicolumn{12}{c|}{Scenario 2 ($d_{OBD} = 3$)} \\
\hline
STEIN & 1.5 & 16.4 & \textbf{70.2} & 11.2 & 0.6 & 0.1 & 5.3 & 13.1 & \textbf{22.1} & 4.1 & 0.3 & 55.0 \\ 
  TITE-BOIN12 & 4.2 & 18.4 & \textbf{54.4} & 20.9 & 1.9 & 0.2 & 6.5 & 12.3 & \textbf{18.2} & 6.6 & 1.3 & 26.3 \\ 
  TITE-BOIN-ET & 2.2 & 15.1 & \textbf{59.9} & 20.6 & 2.1 & 0.1 & 4.8 & 12.3 & \textbf{21.0} & 5.9 & 0.9 & 25.3 \\ 
  TITE-STEIN & 1.7 & 21.9 & \textbf{67.0} & 9.1 & 0.2 & 0.1 & 6.4 & 14.1 & \textbf{20.6} & 3.7 & 0.1 & 25.4 \\ 
  LO-TC & 0.0 & 0.6 & \textbf{22.6} & 39.1 & 37.7 & 0.0 & 3.5 & 4.3 & \textbf{10.5} & 13.4 & 13.3 & 27.3 \\ 
  Joint TITE-CRM & 0.1 & 2.2 & \textbf{23.4} & 56.6 & 16.7 & 1.0 & 3.8 & 7.0 & \textbf{12.0} & 14.9 & 7.0 & 16.8 \\ 
\hline
Design & \multicolumn{12}{c|}{Scenario 3 ($d_{OBD} = 4$)} \\
\hline
STEIN & 0.4 & 3.0 & 18.2 & \textbf{65.3} & 12.4 & 0.7 & 4.3 & 5.6 & 8.7 & \textbf{16.8} & 9.4 & 54.8 \\  
  TITE-BOIN12 & 1.8 & 5.9 & 23.4 & \textbf{54.7} & 14.0 & 0.2 & 4.3 & 6.0 & 10.3 & \textbf{16.8} & 7.5 & 30.7 \\ 
  TITE-BOIN-ET & 1.5 & 0.7 & 12.3 & \textbf{52.1} & 32.8 & 0.6 & 4.6 & 5.9 & 8.1 & \textbf{16.7} & 9.6 & 30.9 \\ 
  TITE-STEIN & 0.2 & 1.3 & 18.4 & \textbf{66.9} & 12.9 & 0.3 & 4.5 & 5.7 & 9.4 & \textbf{17.2} & 8.2 & 30.5 \\ 
  LO-TC & 0.0 & 0.0 & 0.6 & \textbf{13.1} & 86.0 & 0.3 & 3.0 & 3.0 & 3.3 & \textbf{5.9} & 29.7 & 29.2 \\ 
  Joint TITE-CRM & 0.1 & 0.0 & 0.9 & \textbf{29.0} & 67.2 & 2.8 & 3.5 & 4.1 & 4.6 & \textbf{13.9} & 18.1 & 16.7 \\  
\hline
Design & \multicolumn{12}{c|}{Scenario 4 (No OBD)} \\
\hline
STEIN & 12.7 & 18.6 & 23.0 & 4.0 & 0.3 & \textbf{41.4} & 9.8 & 13.2 & 12.7 & 3.0 & 0.3 & 47.5 \\ 
  TITE-BOIN12 & 15.8 & 18.6 & 24.1 & 3.8 & 0.5 & \textbf{37.2} & 11.1 & 12.6 & 10.7 & 4.4 & 1.0 & 29.4 \\ 
  TITE-BOIN-ET & 7.3 & 16.4 & 35.0 & 5.1 & 0.4 & \textbf{35.8} & 8.7 & 12.4 & 12.3 & 5.2 & 1.2 & 29.2 \\ 
  TITE-STEIN & 5.6 & 17.9 & 15.4 & 4.3 & 0.2 & \textbf{56.6} & 11.3 & 15.0 & 12.6 & 2.9 & 0.2 & 29.1 \\ 
  LO-TC & 0.1 & 4.5 & 42.0 & 14.8 & 5.1 & \textbf{33.5} & 3.3 & 5.5 & 14.2 & 8.4 & 8.6 & 26.7 \\ 
  Joint TITE-CRM & 0.4 & 22.7 & 17.1 & 0.8 & 0.0 & \textbf{59.0} & 4.0 & 12.4 & 10.4 & 2.8 & 0.6 & 11.9 \\ 
\hline
Design & \multicolumn{12}{c|}{Scenario 5 ($d_{OBD} = 2$)} \\
\hline
STEIN & 10.5 & \textbf{69.6} & 15.2 & 4.7 & 0.0 & 0.0 & 10.6 & \textbf{25.3} & 5.5 & 2.8 & 0.9 & 55.0 \\ 
  TITE-BOIN12 & 16.0 & \textbf{67.8} & 12.8 & 3.4 & 0.0 & 0.0 & 11.9 & \textbf{23.6} & 6.7 & 2.2 & 0.5 & 24.8 \\ 
  TITE-BOIN-ET & 18.3 & \textbf{64.1} & 13.7 & 2.6 & 0.5 & 0.8 & 11.6 & \textbf{22.7} & 5.9 & 3.1 & 1.6 & 24.4 \\ 
  TITE-STEIN & 12.3 & \textbf{68.7} & 14.6 & 4.3 & 0.1 & 0.0 & 11.8 & \textbf{25.2} & 5.6 & 1.8 & 0.5 & 23.8 \\ 
  LO-TC & 0.5 & \textbf{18.4} & 51.2 & 25.1 & 4.4 & 0.4 & 4.6 & \textbf{8.4} & 14.0 & 9.3 & 8.5 & 28.8 \\ 
  Joint TITE-CRM & 0.0 & \textbf{0.1} & 0.8 & 18.2 & 80.8 & 0.1 & 3.2 & \textbf{4.2} & 5.3 & 10.9 & 21.4 & 16.9 \\ 
\hline
Design & \multicolumn{12}{c|}{Scenario 6 ($d_{OBD} = 3$)} \\
\hline
STEIN & 1.6 & 19.9 & \textbf{63.6} & 13.4 & 1.2 & 0.3 & 4.9 & 10.5 & \textbf{21.3} & 6.9 & 1.3 & 54.9 \\ 
  TITE-BOIN12 & 3.9 & 21.0 & \textbf{61.8} & 11.0 & 2.1 & 0.2 & 5.9 & 12.5 & \textbf{18.5} & 6.3 & 1.7 & 27.9 \\ 
  TITE-BOIN-ET & 2.9 & 9.5 & \textbf{64.0} & 20.6 & 2.7 & 0.3 & 4.8 & 9.3 & \textbf{20.3} & 8.1 & 2.5 & 27.7 \\ 
  TITE-STEIN & 2.1 & 21.8 & \textbf{63.8} & 11.0 & 1.0 & 0.3 & 5.5 & 11.6 & \textbf{21.1} & 6.0 & 0.8 & 27.3 \\ 
  LO-TC & 0.1 & 3.0 & \textbf{26.0} & 41.5 & 29.4 & 0.0 & 3.2 & 4.0 & \textbf{9.8} & 11.7 & 16.3 & 28.8 \\ 
  Joint TITE-CRM & 0.0 & 4.0 & \textbf{34.8} & 44.6 & 14.7 & 1.9 & 3.7 & 8.1 & \textbf{12.7} & 12.8 & 7.1 & 16.7 \\ 
\hline
Design & \multicolumn{12}{c|}{Scenario 7 ($d_{OBD} = 4$)} \\
\hline
STEIN & 2.2 & 10.0 & 39.2 & \textbf{43.0} & 4.4 & 1.2 & 5.3 & 8.6 & \textbf{14.4} & 14.1 & 2.5 & 54.7 \\ 
  TITE-BOIN12 & 3.8 & 11.2 & 35.6 & \textbf{42.0} & 6.6 & 0.8 & 5.9 & 8.6 & 13.7 & \textbf{12.7} & 3.9 & 28.9 \\ 
  TITE-BOIN-ET & 2.1 & 4.3 & 32.2 & \textbf{50.0} & 10.8 & 0.6 & 4.7 & 7.1 & 14.1 & \textbf{14.8} & 4.1 & 28.5 \\ 
  TITE-STEIN & 2.3 & 11.8 & 44.2 & \textbf{37.5} & 3.5 & 0.7 & 5.7 & 9.4 & 16.2 & \textbf{11.8} & 1.8 & 28.3 \\ 
  LO-TC & 0.1 & 0.2 & 3.6 & \textbf{23.1} & 73.0 & 0.0 & 3.1 & 3.4 & 4.7 & \textbf{8.5} & 25.3 & 28.7 \\ 
  Joint TITE-CRM & 0.0 & 3.9 & 18.3 & \textbf{40.5} & 33.5 & 3.8 & 3.8 & 7.3 & 9.7 & \textbf{12.1} & 10.8 & 16.5 \\ 
\hline
Design & \multicolumn{12}{c|}{Scenario 8 (No OBD)} \\
\hline
STEIN & 19.0 & 7.3 & 1.0 & 0.1 & 0.0 & \textbf{72.6} & 18.1 & 6.7 & 1.0 & 0.1 & 0.0 & 31.7 \\ 
  TITE-BOIN12 & 28.5 & 6.9 & 0.6 & 0.0 & 0.0 & \textbf{64.0} & 19.5 & 8.1 & 2.6 & 0.5 & 0.1 & 20.5 \\ 
  TITE-BOIN-ET & 11.0 & 7.5 & 0.9 & 0.0 & 0.0 & \textbf{80.6} & 15.2 & 11.1 & 3.6 & 0.7 & 0.1 & 20.7 \\ 
  TITE-STEIN & 8.7 & 8.9 & 2.0 & 0.0 & 0.0 & \textbf{80.4} & 19.8 & 8.2 & 1.2 & 0.1 & 0.0 & 18.7 \\ 
  LO-TC & 8.8 & 25.5 & 4.0 & 0.3 & 0.2 & \textbf{61.2} & 8.3 & 13.1 & 7.1 & 2.5 & 1.6 & 20.7 \\ 
  Joint TITE-CRM & 12.5 & 2.3 & 0.7 & 0.1 & 0.0 & \textbf{84.4} & 13.2 & 5.1 & 0.6 & 0.2 & 0.0 & 8.1 \\ 
\hline
Design & \multicolumn{12}{c|}{Scenario 9 ($d_{OBD} = 2$)} \\
\hline
STEIN & 12.0 & \textbf{63.4} & 20.9 & 1.6 & 0.1 & 2.0 & 7.4 & \textbf{21.2} & 13.3 & 2.5 & 0.2 & 54.6 \\  
  TITE-BOIN12 & 14.5 & \textbf{54.4} & 25.4 & 3.8 & 0.2 & 1.7 & 10.3 & \textbf{18.6} & 11.5 & 3.5 & 0.7 & 27.5 \\ 
  TITE-BOIN-ET & 6.6 & \textbf{44.0} & 39.9 & 6.1 & 1.0 & 2.4 & 7.4 & \textbf{17.8} & 13.5 & 4.9 & 1.0 & 27.3 \\ 
  TITE-STEIN & 13.6 & \textbf{63.9} & 18.4 & 2.1 & 0.1 & 1.9 & 9.3 & \textbf{20.9} & 12.2 & 2.2 & 0.2 & 26.5 \\ 
  LO-TC & 1.6 & \textbf{23.1} & 44.8 & 19.6 & 10.4 & 0.5 & 3.8 & \textbf{9.0} & 14.2 & 9.0 & 8.9 & 28.1 \\ 
  Joint TITE-CRM & 2.5 & \textbf{53.0} & 34.5 & 3.0 & 0.1 & 6.9 & 5.3 & \textbf{21.6} & 11.2 & 3.6 & 0.9 & 16.2 \\ 
\hline
Design & \multicolumn{12}{c|}{Scenario 10 ($d_{OBD} = 3$)} \\
\hline
STEIN & 0.5 & 9.0 & \textbf{53.7} & 26.7 & 8.5 & 1.6 & 5.5 & 9.2 & \textbf{18.8} & 8.3 & 2.7 & 54.4 \\ 
  TITE-BOIN12 & 2.5 & 12.4 & \textbf{48.9} & 24.4 & 11.1 & 0.7 & 5.5 & 9.6 & \textbf{17.0} & 8.7 & 4.0 & 27.8 \\ 
  TITE-BOIN-ET & 1.6 & 6.3 & \textbf{45.0} & 31.3 & 15.3 & 0.5 & 4.5 & 7.9 & \textbf{17.9} & 9.5 & 5.1 & 27.4 \\ 
  TITE-STEIN & 1.2 & 11.2 & \textbf{56.5} & 24.0 & 6.5 & 0.6 & 6.2 & 10.0 & \textbf{18.8} & 7.8 & 2.0 & 27.4 \\ 
  LO-TC & 0.0 & 0.2 & \textbf{2.1} & 10.6 & 87.0 & 0.1 & 3.1 & 3.4 & \textbf{4.6} & 6.1 & 27.7 & 28.6 \\ 
  Joint TITE-CRM & 0.2 & 2.5 & \textbf{5.3} & 19.0 & 69.1 & 3.9 & 4.5 & 6.4 & \textbf{6.2} & 8.7 & 17.8 & 16.5 \\ 
\hline
Design & \multicolumn{12}{c|}{Scenario 11 ($d_{OBD} = 4$)} \\
\hline
STEIN & 1.3 & 9.3 & 22.5 & \textbf{57.3} & 8.5 & 1.1 & 5.0 & 7.6 & 9.9 & \textbf{14.8} & 7.3 & 54.7 \\  
  TITE-BOIN12 & 5.3 & 14.3 & 23.5 & \textbf{45.4} & 10.7 & 0.8 & 5.7 & 8.0 & 10.3 & \textbf{14.1} & 6.7 & 31.5 \\ 
  TITE-BOIN-ET & 3.1 & 3.2 & 18.1 & \textbf{49.8} & 24.9 & 0.9 & 5.9 & 8.4 & 9.6 & \textbf{13.3} & 7.7 & 32.4 \\ 
  TITE-STEIN & 2.4 & 10.3 & 25.9 & \textbf{52.5} & 7.7 & 1.2 & 5.5 & 8.0 & 10.9 & \textbf{14.3 }& 6.3 & 31.3 \\ 
  LO-TC & 0.0 & 0.4 & 1.4 & \textbf{14.5} & 82.9 & 0.8 & 3.1 & 3.2 & 3.7 & \textbf{5.7} & 29.2 & 29.5 \\ 
  Joint TITE-CRM & 0.3 & 0.3 & 16.1 & \textbf{35.7} & 29.9 & 17.7 & 3.6 & 5.5 & 8.7 & \textbf{12.2} & 10.4 & 15.4 \\ 
\hline
Design & \multicolumn{12}{c|}{Scenario 12 (No OBD)} \\
\hline
STEIN & 7.4 & 10.0 & 18.6 & 14.1 & 1.7 & \textbf{48.2} & 8.8 & 10.4 & 10.6 & 6.3 & 1.8 & 46.2 \\ 
  TITE-BOIN12 & 9.7 & 9.5 & 15.7 & 10.1 & 2.6 & \textbf{52.4} & 9.9 & 10.8 & 9.4 & 5.6 & 2.4 & 31.9 \\ 
  TITE-BOIN-ET & 6.0 & 8.2 & 29.2 & 16.9 & 5.2 & \textbf{34.5} & 8.4 & 10.5 & 10.6 & 7.5 & 3.4 & 33.0 \\ 
  TITE-STEIN & 1.1 & 5.9 & 10.1 & 8.6 & 0.9 & \textbf{73.4} & 10.1 & 12.1 & 11.5 & 6.2 & 1.5 & 31.8 \\ 
  LO-TC & 0.0 & 0.0 & 2.2 & 7.6 & 10.8 & \textbf{79.4} & 3.0 & 3.3 & 5.4 & 6.9 & 10.6 & 22.8 \\ 
  Joint TITE-CRM & 0.1 & 1.4 & 10.3 & 7.6 & 1.4 & \textbf{79.2} & 3.9 & 5.7 & 7.2 & 6.4 & 2.2 & 10.3 \\
\hline
\end{longtable}
\begin{tablenotes}
\footnotesize
\setlength{\baselineskip}{0.6\baselineskip}
\item ET: Early termination.
\end{tablenotes}

\clearpage
\begin{table}[]
\caption{ Summary of toxicity and efficacy data by dose level in the TRANSCEND NHL 001 trial}
\label{tab:case_summary}
\centering
\begin{tabular}{|c|ccc|c|ccc|}
\hline
Toxicity Summary & DL1 & DL2 & DL3 & Efficacy Summary & DL1 & DL2 & DL3\\
\hline
DLT evaluable sample size & 45 & 50 & 41 & Efficacy evaluable sample size & 40 & 169 & 41 \\ 
Number of DLTs & 6 & 2 & 1 & Number of responders & 27 & 125 & 30 \\
Observed DLT rate & 13\% & 4\% & 2\% & Observed ORR & 68\% & 74\% & 73\% \\
Assumed DLT rate & 7\% & 10\% & 12\% & Assumed ORR & 65\% & 75\% & 75\% \\
\hline 
\end{tabular}
\end{table}

\begin{table}[]
    \caption{Comparison of TITE-BOIN12, TITE-BOIN-ET, and TITE-STEIN based on the TRANSCEND NHL 001 trial}
\label{tab:case_result}
\centering
\begin{tabular}{|c|cccc|ccc|c|}
\hline
& \multicolumn{4}{c|}{Selection Probability (\%)} & \multicolumn{3}{c|}{Number of Patients} & Duration \\
\hline
Design & DL1 & DL2 & DL3 & ET & DL1 & DL2 & DL3 & (Months)\\
\hline
TITE-BOIN12 & 32.8 & \textbf{44.8} & 18.4 & 0.0 & 55.0 & \textbf{59.4} & 20.6 & 31.1\\
TITE-BOIN-ET & 54.2 & \textbf{37.2} & 8.6 & 0.0 & 76.2 & \textbf{48.3} & 10.5 & 29.8\\
TITE-STEIN & 28.5 & \textbf{55.6} & 15.9 & 0.0 & 51.9 & \textbf{65.5} & 17.6 & 30.7\\ 
\hline 
\end{tabular}
\end{table}

\clearpage

\end{document}